**TYMOSHCHUK Dmytro**
Ternopil Ivan Puluj National Technical University
https://orcid.org/0000-0003-0246-2236
e-mail: dmytro.tymoshchuk@gmail.com

**YATSKIV Vasyl**
West Ukrainian National University
https://orcid.org/0000-0001-9778-6625
e-mail: jazkiv@ukr.net

**TYMOSHCHUK Vitaliy**
Ternopil Ivan Puluj National Technical University
https://orcid.org/0009-0007-2858-9434
e-mail: vitaliy.tymoshchuk@gmail.com

**YATSKIV Nataliya**
West Ukrainian National University
https://orcid.org/0000-0003-2421-4217
e-mail: jatskiv@ukr.net

# INTERACTIVE CYBERSECURITY TRAINING SYSTEM BASED ON SIMULATION ENVIRONMENTS

*Rapid progress in the development of information technology has led to a significant increase in the number and complexity of cyber threats. Traditional methods of cybersecurity training based on theoretical knowledge do not provide a sufficient level of practical skills to effectively counter real threats. The article explores the possibilities of integrating simulation environments into the cybersecurity training process as an effective approach to improving the quality of training. The article presents the architecture of a simulation environment based on a cluster of KVM hypervisors, which allows creating scalable and flexible platforms at minimal cost. The article describes the implementation of various scenarios using open source software tools such as pfSense, OPNsense, Security Onion, Kali Linux, Parrot Security OS, Ubuntu Linux, Oracle Linux, FreeBSD, and others, which create realistic conditions for practical training.*

**ТИМОЩУК Дмитро**
Тернопільський національний технічний університет імені Івана Пулюя

**ЯЦКІВ Василь**
Західноукраїнський національний університет

**ТИМОЩУК Віталій**
Тернопільський національний технічний університет імені Івана Пулюя

**ЯЦКІВ Наталія**
Західноукраїнський національний університет

# ІНТЕРАКТИВНА СИСТЕМА НАВЧАННЯ З КІБЕРБЕЗПЕКИ НА ОСНОВІ СИМУЛЯЦІЙНИХ СЕРЕДОВИЩ

*Швидкий прогрес в розвитку інформаційних технологій спричинив суттєве зростання кількості й складності кіберзагроз. Традиційні методи навчання з кібербезпеки, що базуються на теоретичних знаннях, не забезпечують достатнього рівня практичних навичок для ефективної протидії реальним загрозам. У статті досліджено можливості інтеграції симуляційних середовищ у процес навчання з кібербезпеки як ефективного підходу до підвищення якості підготовки фахівців. Представлено архітектуру симуляційного середовища на базі кластеру гіпервізорів KVM, що дозволяє створювати масштабовані та гнучкі платформи з мінімальними витратами. Описано реалізацію різноманітних сценаріїв з використанням відкритих інструментів, таких як pfSense, OPNsense, Security Onion, Kali Linux, Parrot Security OS, Ubuntu Linux, Oracle Linux, FreeBSD та інших, які створюють реалістичні умови для практичного навчання.*

*Використання симуляційних середовищ у навчанні з кібербезпеки має кілька переваг. По-перше, це можливість працювати з реальними інструментами та технологіями, які використовуються в промисловості. По-друге, симуляції дозволяють адаптувати сценарії до конкретних потреб і рівня підготовки користувачів, забезпечуючи індивідуальний підхід до навчання. По-третє, інтерактивний формат сприяє підвищенню мотивації та залученості учасників, що позитивно впливає на ефективність навчання. Незважаючи на очевидні переваги, впровадження симуляційних середовищ у навчальний процес пов'язане з певними труднощами. До них належать потреба у значних технічних ресурсах для розгортання та підтримки таких систем, а також необхідність постійного оновлення сценаріїв відповідно до нових кіберзагроз. Крім того, важливо забезпечити інтеграцію симуляційних середовищ у загальну навчальну програму, щоб вони доповнювали теоретичні знання та сприяли всебічному розвитку навичок. Особливої актуальності набуває розробка інтерактивних систем навчання, які поєднують переваги симуляційних середовищ з ефективними методами навчання. Такі системи мають бути гнучкими, масштабованими та здатними адаптуватися до потреб різних категорій користувачів – від студентів до досвідчених професіоналів.*

*Метою цієї статті є дослідження можливостей інтеграції симуляційних середовищ у навчальний процес з кібербезпеки та розробка моделі інтерактивної системи навчання, яка сприятиме підвищенню якості навчання в цій галузі.*
*Ключові слова: Гіпервізор, KVM, IDS, IPS, OPNsense, pfSense, Suricata, Snort, Security Onion, Cluster.*

## INTRODUCTION

With the dynamic development of information technology and the growing dependence of businesses, government agencies and individuals on digital systems, the number and complexity of cyber threats are increasing significantly. Attackers are using increasingly sophisticated attack methods aimed at compromising the confidentiality, integrity and availability of information systems. This creates serious challenges for organizations and requires a high level of training for cybersecurity professionals.

Traditional teaching methods based on theoretical knowledge and passive learning do not meet the current requirements for training specialists in this field. They do not provide a sufficient level of practical skills necessary to effectively counter real cyber threats. Lack of practical experience can lead to ineffective security measures and an increased risk of successful attacks. In response to these challenges, interactive training methods using simulation environments are becoming increasingly common. Such environments allow simulating real-life cyber-attack scenarios in a controlled environment, enabling professionals to practice their skills in detecting, analysing and responding to threats without risking real systems. Simulation platforms provide a realistic environment that promotes a deeper understanding of attack mechanisms and effective defence methods.

There are several advantages to using simulation environments in cybersecurity training. Firstly, it is an opportunity to work with real tools and technologies used in industry. Secondly, simulations allow us to adapt scenarios to the specific needs and level of training of users, providing an individual approach to training. Thirdly, the interactive format helps to increase the motivation and engagement of participants, which has a positive impact on the efficiency of learning. Despite the obvious advantages, the introduction of simulation environments into the learning process is associated with certain challenges. These include the need for significant technical resources to deploy and maintain such systems, as well as the need to constantly update scenarios in line with new cyber threats. In addition, it is important to ensure that simulation environments are integrated into the overall curriculum so that they complement theoretical knowledge and promote comprehensive skill development. Of particular relevance is the development of interactive learning systems that combine the advantages of simulation environments with effective teaching methods. Such systems should be flexible, scalable and able to adapt to the needs of different categories of users - from students to experienced professionals.

The purpose of this article is to explore the possibilities of integrating simulation environments into the cybersecurity education process and to develop a model of an interactive learning system that will help improve the quality of training in this area.

## MAIN PART

One of the main advantages of simulation environments is their ability to reproduce realistic conditions of information systems and networks. This is achieved through the use of virtualisation, emulation of network protocols and services, and integration with real-world security tools. This approach allows specialists to practice scenarios with various types of attacks, including DDoS, phishing, SQL injection, social engineering attacks, and others. Modern simulation environments allow not only to model attacks but also to practice defence strategies, configure IDS/IPS, firewalls, and network traffic monitoring and analysis systems. This ensures a comprehensive approach to training, allowing specialists to understand both the tactics and techniques of attackers and the methods of effective defence.

There are a significant number of commercial simulation environments on the market designed to educate and train cybersecurity professionals. One of the most well-known platforms is Cyberbit Range, which provides the ability to recreate complex cyberattack scenarios in a secure environment [1]. It supports a wide range of attacks, including DDoS, phishing, database attacks, and others. Cyberbit Range integrates with real security tools, such as SIEM systems and firewalls, allowing users to work with the same tools they use in their daily activities. The platform also provides the ability to assess team performance, identify knowledge gaps, and improve overall cyber readiness. RangeForce is a cloud-based platform that offers interactive training modules and labs [2]. It allows users to develop practical skills by working with real-world tools and technologies. The platform supports team training, which helps develop communication and leadership skills. RangeForce adapts to the user's level of knowledge, making it suitable for both beginners and experienced professionals. The IBM X-Force Cyber Range focuses on preparing for complex cyber incidents [3]. The platform provides realistic attack scenarios where participants can practice their skills in conditions as close as possible to real crisis situations. Particular emphasis is placed on developing teamwork skills, decision-making under stress, and interaction with different parts of the organisation during an incident. The Cisco Cyber Range offers a comprehensive training programme that includes gradually increasingly complex scenarios and focuses on methodologies and techniques independent of specific tools [4]. This allows professionals to apply the acquired knowledge in different environments and with different technologies. The platform helps to develop technical skills and improve teamwork. Project Ares, developed by Circadence, combines game elements with realistic cyber scenarios [5]. It includes a wide range of training modules, from basic labs to complex missions that simulate real-world attacks. The use of realistic network environments and virtual machines allows users to work with authentic tools and technologies. The International Telecommunication Union (ITU) also uses simulation platforms to conduct cyber exercises and professional development. The Cyber Ranges platform used by the ITU allows simulating realistic cyberattack scenarios and practising incident response skills in a secure environment [6]. This promotes international cooperation and exchange of experience in the field of cybersecurity.

While there are many commercial cybersecurity training simulation environments on the market, they are often expensive. The high cost of licences, support and updates can be a significant barrier for educational institutions and organisations on a budget. This limits access to modern tools and technologies needed to effectively train cybersecurity professionals. Given these challenges, many organisations are turning to alternative solutions that are more cost-effective. The use of open source software and free tools allows for the creation of effective simulation environments with minimal financial outlay. Such solutions not only reduce the cost but also provide flexibility in customisation and adaptation to the specific needs of the training process.

For example, using open-source hypervisors such as Xen and KVM allows us to deploy virtual machines without having to purchase expensive licences [7]. Security Onion, which includes ELK Stack, IDS Suricata, Zeek, and CyberChef, is a free platform for monitoring and analysing network traffic [8]. It provides powerful tools for detecting and responding to cyber threats without requiring significant financial investment. Free firewalls and routers, such as pfSense and OPNsense, provide extensive network security configuration options [9][10]. They support packet filtering, VPNs, load balancing, and Snort and Suricata IDS/IPS, making them suitable for use in simulation environments. The use of MikroTik CHR (Cloud Hosted Router) virtual router allows us to create flexible and scalable network solutions with support for a variety of network protocols and security features [11]. To simulate attacks and practice penetration testing skills, it is possible to use distributions such as Kali Linux and Parrot Security OS, which are free and contain a wide range of cybersecurity tools. To create vulnerable environments that can be used as targets for attacks, Metasploitable VM is used - a virtual machine with deliberately built-in vulnerabilities [12]. It allows us to safely practice exploiting vulnerabilities and test the effectiveness of security mechanisms.

Figure 1 shows the architecture of a simulation environment based on a cluster of KVM hypervisors [13], which is used to create interactive cybersecurity training platforms.

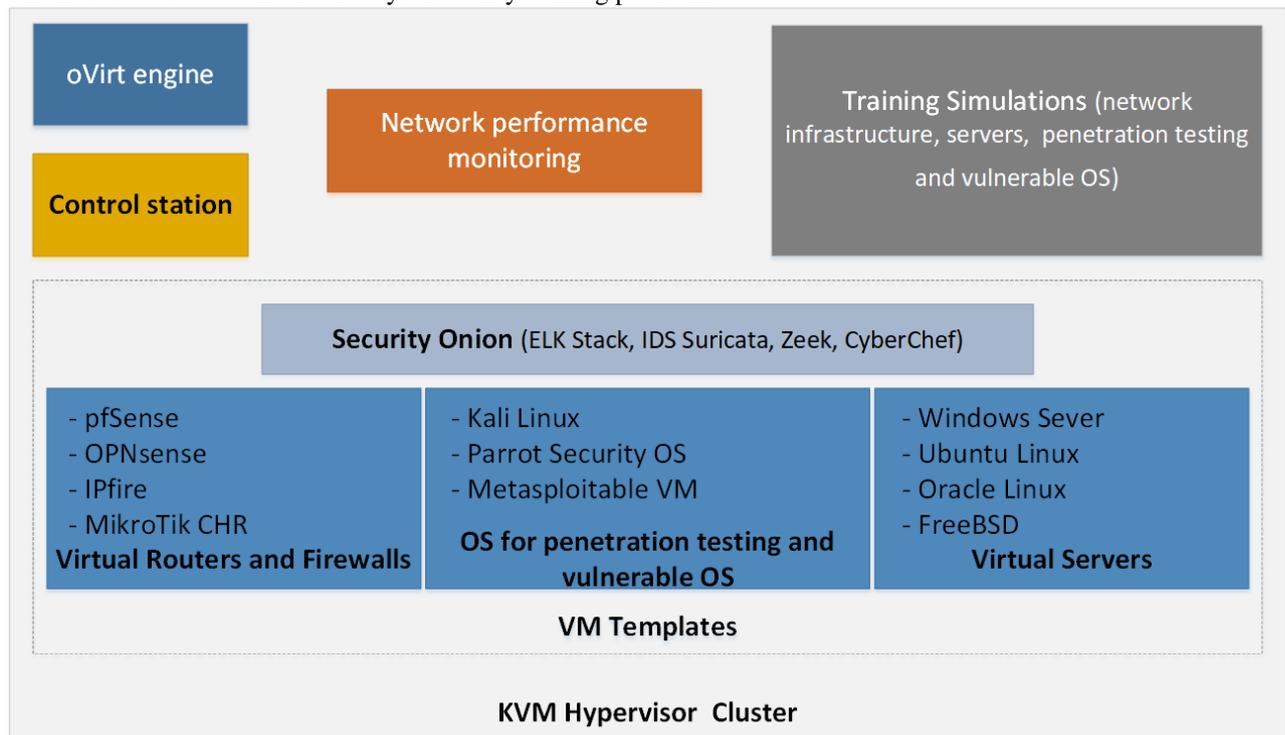

**Fig. 1 Simulation environment architecture**

The basis of this environment is KVM, which provides virtualisation of server resources, allowing efficient deployment of virtual machines for various tasks [14]. For centralised management, the oVirt Engine is used, which provides administration of virtualised resources, including performance monitoring, management of virtual machine templates, and control of infrastructure usage [15]. The control station acts as a centralised control point for the environment. It allows us to administer simulation platforms, monitor the state of the infrastructure, and control its performance. Network performance monitoring allows us to analyse traffic, detect anomalies, prevent network congestion, and maintain the stability of the simulation environment.

Training simulations are pre-configured scenarios that mimic real-world cybersecurity situations for students and professionals. They allow users to learn the basics of networking, security systems, and methods of detecting and countering cyber threats. These simulations are aimed at modelling attack, incident and defence scenarios to improve participants' skills and prepare them for real-world challenges. The ready-made scenarios cover various aspects of cyber defence, such as detecting malicious traffic, countering DDoS attacks, analysing logs, configuring firewalls and penetration testing. In these environments, users work with predefined tools and platforms that simulate real-world conditions. For example, they can deploy virtual routers and firewalls based on pfSense or OPNsense, examine IDS and IPS performance with Suricata or Snort, and analyse event logs with ELK Stack in Security Onion. Integrated specialised vulnerable systems like Metasploitable VM allow users to explore vulnerabilities and test defence mechanisms. Scenarios also include real-world threat elements, such as using Kali Linux or Parrot Security OS to perform ethical hacking and test systems for vulnerabilities.

The platform supports the deployment of virtual servers based on popular operating systems such as Windows Server, Ubuntu Linux, Oracle Linux and FreeBSD. This allows users to create a complete infrastructure for training, research, and testing.

This architecture provides the ability to build scalable simulation environments at minimal cost, which allows us to create realistic scenarios for practising cybersecurity skills. All of this makes it possible to provide high flexibility in setting up the environment and support modern requirements for interactive learning.

Figures 2-4 show an example of three basic scenarios implemented in the simulation environment. Each scenario reflects unique configurations that allow users to simulate different situations and approaches to protecting networks and detecting and countering cyber threats.

The first scenario (Figure 2) shows a network where pfSense with Suricata IDS is used to monitor traffic passing through an external virtual switch.

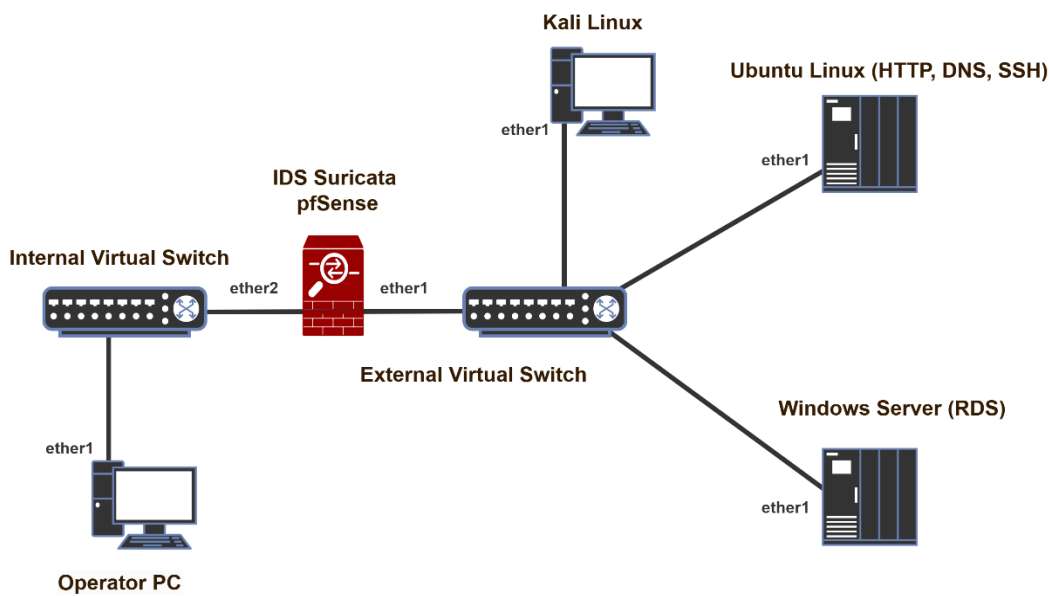

**Fig. 2 Scenario 1 based on pfSense and IDS Suricata**

The Kali Linux operating system is used for pentesting, Ubuntu Linux with configured HTTP, DNS, and SSH services and Windows Server with configured RDS are used as targets for the attack.

The second scenario (Figure 3) involves the use of OPNsense with Snort IPS configured to detect and prevent threats from the external network.

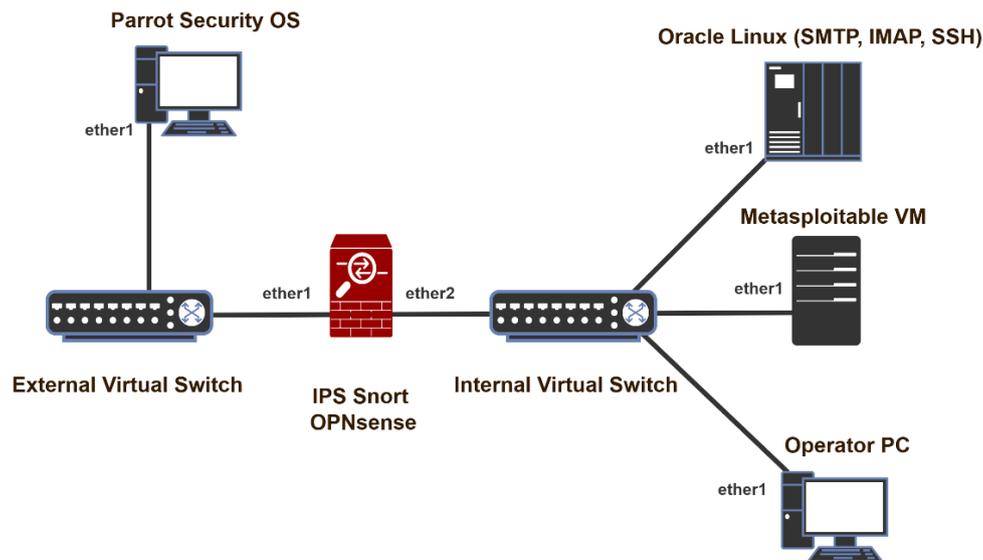

**Fig. 3 Scenario 2 based on OPNsense and Snort IPS**

Parrot Security OS is used to perform penetration tests, Metasploitable VM as a vulnerable target for training, and Oracle Linux to simulate a server with SMTP, IMAP, and SSH support.

The third scenario (Figure 4) demonstrates a more complex infrastructure using Security Onion, which includes components such as ELK Stack, IDS Suricata, Zeek, and CyberChef for in-depth traffic and event analysis.

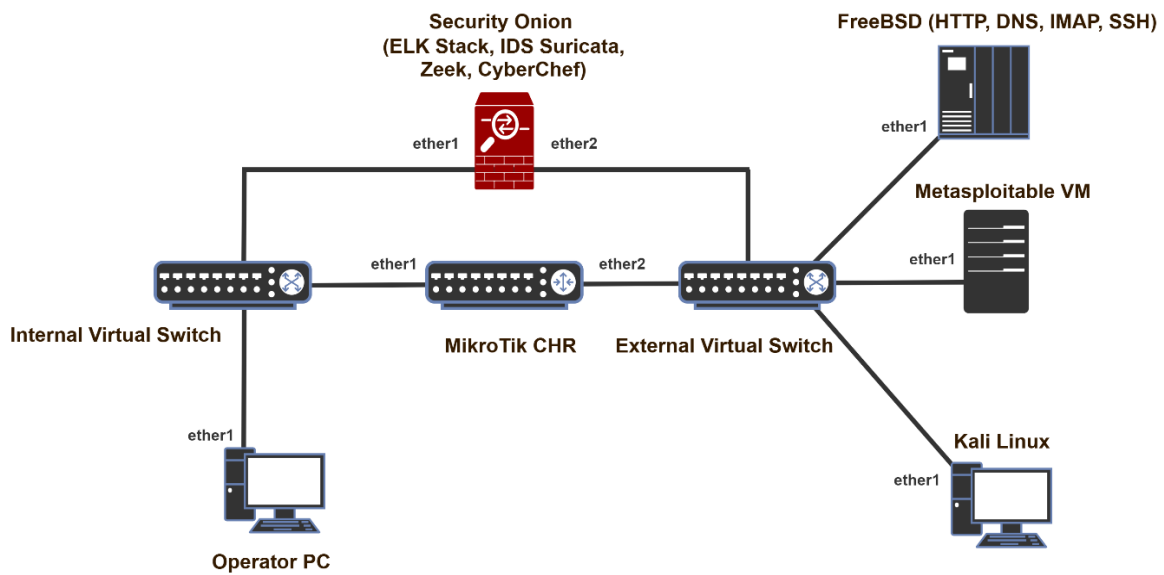

**Fig. 4 Scenario 3 based on Security Onion**

In addition, the network includes MikroTik CHR for routing, Metasploitable VM as a vulnerable server, FreeBSD for simulating different server roles, and Kali Linux for attacks, as well as an operator PC for management.

Each of these scenarios demonstrates the flexibility and scalability of using different technologies for cybersecurity training, testing, and research. They provide a realistic environment for training, improving skills, and testing the effectiveness of defences against real-world threats. Training simulations not only help to gain practical experience in modelling real-life situations, but also provide an opportunity to explore complex scenarios, study modern threats and develop effective methods to counter them.

## CONCLUSIONS

The integration of simulation environments into cybersecurity training is an effective approach to improving the quality of training. It allows combining theoretical knowledge with practical experience, developing the necessary skills and preparing specialists for real challenges in the field of cybersecurity. The presented architecture based on a cluster of KVM hypervisors demonstrates the ability to build scalable and flexible platforms at minimal cost. The implementation of various scenarios using open source tools such as pfSense, OPNsense, Security Onion, Kali Linux, and others allows for realistic conditions for practical training. The use of open source software makes such solutions accessible and cost-effective, contributing to the wider adoption of innovative teaching methods.